\documentclass[12pt]{article}
\usepackage[mathscr]{eucal}
\usepackage{amssymb,latexsym}
\usepackage{verbatim}
\usepackage{amsmath}
\usepackage{amsthm}
\usepackage{enumerate}
\usepackage[T1]{fontenc}
\usepackage[utf8]{inputenc}
\usepackage{authblk}

\newtheorem{thm}{Theorem}[section]
\newtheorem{lem}[thm]{Lemma}
\newtheorem{Def}{Definition}[section]
\newtheorem{prop}[thm]{Proposition}
\newtheorem{rem}[Def]{Remark}

\newcommand\calT{{\mathcal{T}}}

\newcommand\Th{\Theta}

\renewcommand\l{\lambda}

\newcommand\bbR{{\mathbb R}}

\newcommand\vep{\varepsilon}

\renewcommand\S{\Sigma}

\renewcommand\d{\partial}

\newcommand\f{\phi}
\renewcommand\L{\triangle}
\newcommand\D{\nabla}
\newcommand\e{\epsilon}

\renewcommand\div{{\rm div}}

\newcommand\<{\langle}
\renewcommand\>{\rangle}

\renewcommand\k{\kappa}

\renewcommand\l{\lambda}

\newcommand\8{\infty}
\renewcommand\a{\alpha}

\renewcommand\th{\theta}
\renewcommand\Th{\Theta}

\newcommand\beq{\begin{equation}}
\newcommand\eeq{\end{equation}}
\newcommand\ben{\begin{enumerate}}
\newcommand\een{\end{enumerate}}
\newcommand\bit{\begin{itemize}}
\newcommand\eit{\end{itemize}}

\DeclareMathOperator{\diver}{div}

\renewcommand{\div}{\diver}

\newcommand{\tr}{\mathrm{tr}\,}

\newcommand{\R}{\mathbb R}

\newcounter{mnotecount}

\title{Rigidity of marginally outer trapped $2$-spheres}

\author[1]{Gregory J. Galloway}
\author[2]{Abra\~ao Mendes}

\affil[1]{Department of Mathematics, 

University of Miami, Coral Gables, FL, USA}

\affil[2]{Instituto de Matem\'atica, 

Universidade Federal de Alagoas, Macei\'o,  AL, Brazil}

\begin{document}
\date{}
\maketitle
\vspace{.2in}

\begin{abstract} 
In a matter-filled spacetime, perhaps with positive cosmological constant, a stable marginally outer trapped $2$-sphere  must satisfy a certain area inequality.  Namely, as discussed in the paper, its area must be bounded above by $4\pi/c$, where $c > 0$ is a  lower bound on a natural energy-momentum term.  We then consider the rigidity that results for stable, or weakly outermost, marginally outer trapped $2$-spheres that achieve this upper bound on the area.   In particular, we prove a splitting result for $3$-dimensional initial data sets analogous to a result of Bray, Brendle and Neves \cite{Bray} concerning area minimizing $2$-spheres in Riemannian $3$-manifolds with positive scalar curvature.   We further show that these initial data sets locally embed as spacelike hypersurfaces into the Nariai spacetime.  Connections to the Vaidya spacetime and dynamical horizons are also discussed. 
\end{abstract}


\section{Introduction}  

As motivation for the present work, we begin by recalling the following seminal result of Schoen and Yau \cite{sysc} concerning topological obstructions to manifolds of positive scalar curvature.

\begin{thm}[\cite{sysc}]\label{pi1}
Let $(M,g)$ be a closed orientable $3$-manifold of positive scalar curvature, $S > 0$.  Then 
$\pi_1(M)$ does not contain a subgroup isomorphic to that of a surface of genus $g \ge 1$.
\end{thm}

When taken in conjunction with the positive resolution of the surface subgroup conjecture, Theorem \ref{pi1} completely determines the topology of $M$.
The proof combines an area minimization result, together with the following key observation.

\begin{prop}\label{stablemin}
Let $(M,g)$ be an orientable $3$-manifold of positive scalar curvature, $S > 0$.  Then $(M,g)$ does not contain a compact orientable stable minimal surface of positive genus.
\end{prop}

The proof is an application of the formula for the  second variation of area, rewritten in a certain manner; see also  \cite{Gibbons72}.   Both  Propositon \ref{stablemin} and Theorem~\ref{pi1} were generalized by Schoen and Yau to higher dimensions \cite{sysc2}, but in this paper we shall restrict attention to three (spatial) dimensions.   

If, in Proposition \ref{stablemin}, one relaxes the scalar curvature condition to that of nonnegative scalar curvature, $S \ge 0$, then one obtains an {\it infinitesimal} rigidity statement.  As observed in Fischer-Colbrie and Schoen \cite{FCS}, if $\S$ is a compact orientable stable minimal surface of genus $\ge 1$ in a closed orientable $3$-manifold of nonnegative scalar curvature then $\S$ must be a flat totally geodesic torus, with $S = 0$ along $\S$.   Fischer-Colbrie and Schoen also posed the problem of establishing a stronger (more global) rigidity statement, if, say, the torus is suitably area minimizing; cf, \cite[Remark 4]{FCS}.  Cai and the first author \cite{CG} addressed this problem a number of years later, partly motivated by some issues concerning the topology of black holes. 

\begin{thm}[\cite{CG}]\label{theo.CaiGalloway}
Let $(M,g)$ be a $3$-manifold with nonnegative scalar curvature, $S \ge 0$.
If $\S$ is a two-sided torus which is locally area minimizing  then a neighborhood $U$ of $\S$ splits, i.e., $(U,g|_U)$ is isometric to $((-\e,\e)\times \S, dt^2 + h)$, where $h$, the induced metric on $\S$, is flat. 
 \end{thm}
\noindent
It was further shown that if $M$ is complete and $\Sigma$ has least area in its isotopy class, then $M$ is globally flat.   A higher dimensional version of Theorem \ref{theo.CaiGalloway} was obtained in \cite{Cai}; see also \cite{Gschoen} for a simplified proof.  

The torus splitting result in \cite{CG} has  been followed more recently by a number of related rigidity results in three dimensions under different assumptions on the ambient scalar curvature and the topology of the minimal surface; see e.g. \cite{Bray, Nunes, MicallefMoraru, Ambrozio}.  Here we wish to focus on the  result of Bray, Brendle and Neves \cite{Bray}, which we paraphrase as follows.

\begin{thm}[\cite{Bray}]\label{theo.BrayBrendleNeves}
Let $M$ be a Riemannian 3-manifold with scalar curvature bounded from below by $2c$, where $c>0$. If $\Sigma$ is a 2-sphere in $M$ which
is locally area minimizing, then  the area of $\Sigma$ satisfies,
\beq\label{ineq0}
A(\S) \le \frac{4\pi}{c} \,.
\eeq
Moreover, if equality holds then a neighborhood $U$ of $\S$ splits, i.e., 
$(U,g|_U)$ is isometric to $((-\e,\e)\times \S, dt^2 + h)$,  where $h$ is the round metric of radius $1/\sqrt{c}$.
\end{thm}

In fact the area inequality, which also appears in the work of Shen and Zhu~\cite{shen}, only requires $\S$ to be stable.   See e.g. \cite{Gibbons99, ChrisYau, YangYau} for some related inequalities.  A global splitting statement is also obtained in \cite{Bray}.  (For results concerning the regidity of {\it noncompact} area minimizing surfaces see \cite{ChodEich} and references therein.)

From the point of view of relativity, the Bray-Brendle-Neves results may be viewed as statements about time-symmetric (totally geodesic) initial data sets.  The aim of the present paper is to obtain versions of their results for general (non-time-symmetric) initial data sets.  In this more general situation, minimal surfaces are replaced by {\it marginally outer trapped surfaces} (MOTSs); see Section 2 for relevant definitions and properties of MOTSs.
In Section 3, 
we  present an {\it infinitesimal} rigidity result (Proposition \ref{infrigid}), for MOTSs $\S$ which   saturate an area inequality analogous to \eqref{ineq0}.  This is  then used to prove a splitting theorem (Theorem \ref{split}), which is a spacetime analogue of Theorem \ref{theo.BrayBrendleNeves}.  Theorem \ref{split} also bears some relation to the spacetime rigidity result obtained in \cite{Grigid}. 
Some connections to Vaidya spacetime (and dynamical horizons \cite{AK}) and Nariai spacetime are also considered.  In Section 4, it is shown that an outer neighborhood of a MOTS $\S$ which saturates the relevant area inequality, can be realized as a spacelike hypersurface in Nariai spacetime (cf. Theorem \ref{nariai}), thereby locally classifying the geometry of such initial data sets.

\smallskip
\noindent
{\bf Acknowledgments.} 
The work of GJG was partially supported  by NSF grant DMS-1313724.  The work of AM  was carried out while he was a Visiting Graduate Student at Princeton University during the 2015-2016 academic year. AM was partially supported by NSF grant DMS-1104592 and by the CAPES Foundation, Ministry of Education of Brazil.  AM would like to express his gratitude to his Ph.D. advisors Fernando Cod\'a Marques, at Princeton University, and Marcos Petr\'ucio Cavalcante, at UFAL.

\section{Preliminaries}

A marginally outer trapped surface (MOTS) in spacetime represents an extreme gravitational situation: Under suitable circumstances, the occurrence of a MOTS  signals the presence of a black hole \cite{HE,CGS}.   For this and other reasons MOTSs have played  a fundamental role in quasi-local descriptions of  black holes; see e.g.,  \cite{AK}.  MOTSs arose in a more purely mathematical context  in the work of Schoen and Yau \cite{SY2} concerning the existence of solutions of Jang's equation, in connection with their proof of the positive mass theorem.  The mathematical theory of MOTSs has been greatly developed in recent years.  We refer the reader to the survey article \cite{AEM} which describes many of these developments.
  
In this section we recall some basic definitions and facts about MOTSs.  Let $(\bar M,\bar g)$ be a $4$-dimensional spacetime (time oriented Lorentzian manifold).  Consider an initial data set $(M,g,K)$ in  $(\bar M,\bar g)$.  Hence, $M$ is a spacelike hypersurface (of dimension three), and $g$ and $K$ are the induced metric and second fundamental  form, respectively, of $M$.  
To set sign conventions, for vectors $X,Y \in T_pM$, $K$ is defined as, $K(X,Y) = \bar g(\bar \D_X u,Y)$, where $\bar\D$ is the Levi-Civita connection of $\bar M$ and $u$ is the future directed timelike unit normal vector field to $M$. 

Let $\S$ be a closed (compact without boundary) two-sided surface in $M$.   Then $\S$ admits a smooth unit normal field $\nu$ in $M$, unique up to sign.  By convention, refer to such a choice as outward pointing. 
Then $l_+ = u+\nu$ and $l_- =  u - \nu$ are future directed outward pointing and inward pointing, respectively,  null normal vector fields along $\S$.  The null second fundamental forms $\chi_+$ and $\chi_-$ of $\Sigma$ in $\bar M$ are defined by 
\beq\label{nullforms}
\chi_\pm(X,Y)=\bar g(\bar\nabla_Xl_\pm,Y)=K(X,Y)\pm A(X,Y),\ \ \ X,Y\in T_p\Sigma \,,
\eeq
where $A$ is the second fundamental form of $\Sigma$ in $M$.

The {\em null mean curvatures} (or {\em null expansion scalars}) $\theta_\pm$ of $\Sigma$ are obtained by tracing $\chi_\pm$ with respect to 
$\Sigma$,
\beq
\theta_\pm=\tr_\Sigma\chi_\pm=\tr_\Sigma K\pm H \,,
\eeq 
where $H$ is the mean curvature of $\Sigma$ in $M$. In particular, when $M$ is time-symmetric ($K=0$), $\theta_+$ is just 
the mean curvature of $\Sigma$ in $M$. 

As first defined by Penrose, $\S$ is said to be a trapped surface if both $\th_-$ and $\th_+$ are negative.   
Focusing attention on the outward null normal, we say that  $\S$ is an outer trapped surface  if $\th_+ < 0$.  Finally, we define $\S$ to be a marginally
outer trapped surface (MOTS) if $\th_+$ vanishes identically.   Note that in the 
time-symmetric case, a MOTS is just a minimal surface.

\smallskip
Henceforth, to simplify notation, we drop the plus sign, and denote $\theta=\theta_+$, $\chi=\chi_+$, and $l=l_+$.
\medskip

In  \cite{AMS1,AMS2}, Andersson, Mars and Simon introduced a notion of stability for MOTSs, analogous in a certain sense to that for minimal surfaces, which we now recall.   

Let $\S$ be a MOTS in the initial data set $(M,g,K)$ with outward unit normal $\nu$.  We consider a normal variation of $\S$ in $M$,  i.e.,  a variation 
$t \to \S_t$ of $\S = \S_0$ with variation vector field 
$V = \frac{\d}{\d t}|_{t=0} = \phi\nu$,  $\phi \in C^{\infty}(\S)$.
Let $\th(t)$ denote the null expansion of $\S_t$
with respect to $l_t = u + \nu_t$, where $u$ is the future
directed timelike unit normal to $M$ and $\nu_t$ is the
outer unit normal  to $\S_t$ in $M$.   A computation as in \cite{AMS2} gives,
\beq\label{thder} 
\left . \frac{\d\th}{\d t} \right |_{t=0}   =
L(\f) \;, 
\eeq 
where $L : C^{\infty}(\S) \to C^{\infty}(\S)$ is the operator, 
\beq\label{stabop}
L(\phi)  = -\triangle \phi + 2\<X,\D\phi\>  + \left(Q +{\rm div}\, X - |X|^2 \right)\phi \,,
\eeq 
and where,
\beq\label{Q}
Q =  \frac12 S_{\S} - (\mu + J(\nu)) - \frac12 |\chi|^2\,,
\eeq
$\triangle$, $\D$ and ${\rm
div}$ are the Laplacian, gradient and divergence operators,
respectively, on $\S$, $S_{\S}$ is the scalar curvature of $\S$ with respect to the induced metric 
$\<\,,\,\>$ on $\S$, $X$ is the 
vector field  on $\S$  dual to the one form $K(\nu,\cdot)|_{T\S}$, 
and $\mu$ and $J$ are defined in terms
of the Einstein tensor $G = {\rm Ric}_{\bar M} - \frac12 R_{\bar M} \bar g$ by $\mu = G(u,u)$,
$J = G(u,\cdot)$.  When the Einstein equations are assumed to hold, $\mu$ and $J$ represent the  energy density and linear momentum density along $M$.  As a consequence of the Gauss-Codazzi equations, the quantities $\mu$ and $J$ can be expressed solely in terms of initial
data, 
\begin{align*}
\mu &= \frac12\left(R + ({\rm tr}\,K)^2 - |K|^2 \right) \,,  \\
J &= \div K- d({\rm tr}\, K) \,, 
\end{align*} 
where $R$ is the scalar curvature of $M$. 

The operator $L$ is not self-adjoint in general, but does have the following properties; see \cite{AMS2} and references therein.

\begin{lem}\label{prin}  The following holds for the operator $L$.
\ben[(i)]
\item  There is a real eigenvalue $\l_1 = \l_1(L)$, called the principal eigenvalue, such that for any other eigenvalue $\mu$, ${\rm Re}(\mu) \ge \l_1$.    The
associated eigenfunction $\f$, $L(\f) = \l_1 \f$, is unique up to a multiplicative constant, and can be chosen to be strictly positive.
\item $\l_1 \ge 0$ (resp., $\l_1 > 0$) if and only if there exists $\psi \in C^{\infty}(\S)$, $\psi> 0$, such that $L(\psi) \ge 0$ (resp., $L(\psi) > 0$). 
\een
\end{lem}

Our main results will rely on the following key fact.   Consider the 
``symmetrized'' operator
$L_0: C^{\infty}(\S) \to C^{\infty}(\S)$,
\beq\label{stabop0}
L_0(\phi)  = -\triangle \phi  + Q  \phi \,,
\eeq
obtained formally from \eqref{stabop}  by  setting $X= 0$.

\begin{lem}\label{keyfact}
$\l_1(L_0) \ge \l_1(L)$.  Hence, if  $\l_1(L) \ge 0$,
\beq\label{stabineq}
\int_{\S} |\D f|^2 + Q f^2 dA \ge 0 \,,
\eeq
for all $f \in C^{\infty}(\S)$.
\end{lem}

The assertion $\l_1(L_0) \ge \l_1(L)$ follows from the main argument in \cite{GS}; see also \cite{AMS2}, \cite{Gschoen}.   The inequality \eqref{stabineq}  then follows from the Rayleigh formula characterizing the principal eigenvalue of the operator $L_0$,

\beq\label{ray}
\l_1(L_0) = \inf_{f \not\equiv 0} \frac{\int_{\S} |\D f|^2 + Q f^2 \,dA}{\int_{\S} f^2 \,dA}  \,.
\eeq
An inequality similar to \eqref{stabineq} has been obtained in \cite{Dain1}. 

\smallskip
Observe that in the time-symmetric 
case, $L$ reduces to  the classical stability operator for minimal surfaces.  As such, we refer to $L$ 
as the MOTS stability operator associated with variations in the null expansion $\th$.  
In analogy with the minimal surface case, we
say that a MOTS is stable provided $\l_1(L) \ge 0$.  (In the minimal surface case
this is equivalent to the second variation of area being nonnegative.) 

Heuristically, a MOTS $\S$ is stable if it is {\it infinitesimally outermost}.   Stable MOTSs arise in various
situations.   For example, weakly outermost MOTSs are stable.  Indeed, if $\l_1(L) < 0$,  \eqref{thder} implies that $\S$ can be deformed outward to an outer trapped surface.
Weakly outermost MOTSs include, in particular, compact cross sections of the event horizon in stationary black hole spacetimes obeying the null energy condition.  More generally, results of Andersson and
Metzger \cite{AM1, AM2}, and of Eichmair \cite{Eich1, Eich2} establish natural criteria for the existence of 
outermost (and hence weakly outermost) MOTSs; see also \cite{AEM}.

\section{Rigidity results}

\subsection{Infinitesimal rigidity}
We first establish the following infinitesimal rigidity result.   

\begin{prop}\label{infrigid}
Let   $\S$ be a stable spherical (topologically $S^2$) MOTS in a $3$-dimensional initial data $(M,g,K)$.  Suppose there exists $c > 0$, such that  $\mu +J(\nu) \ge c$ on $\S$, where $\nu$ is the outward unit normal to $\S$.   Then the area of $\S$ satisfies,
\beq\label{ineqA}
A(\S) \le \frac{4\pi}{c} \,.
\eeq
Moreover, if equality holds, $\S$ is a round $2$-sphere, with  Gaussian curvature $\kappa_\S = c$, the outward null second fundamental form $\chi$  of $\S$ vanishes, and $\mu +J(\nu) = c$ on 
$\S$.
\end{prop}  

An inequality closely related to \eqref{ineqA} has been obtained by Hayward \cite{Hay1} for spacetimes with positive cosmological constant, in which stability is expressed in terms of variations of the null expansion along a null hypersurface associated with a double null foliation.

\smallskip
In the presence of matter fields and/or a positive cosmological constant,  a positive lower bound on $\mu +J(\nu)$ like that assumed in Proposition \ref{infrigid} is expected.   Indeed, suppose the initial data set $(M,g,K)$ comes from a spacetime $(\bar M, \bar g)$ which satisfies the Einstein equation,
\beq\label{eineq}
G + \Lambda \bar g = \cal{T}
\eeq
where, as in Section 2,  $G = {\rm Ric}_{\bar M} - \frac12 R_{\bar M} \bar g$ is the Einstein tensor, and $\cal T$ is the energy-momentum tensor.   Then, setting $\ell = u+\nu$, where $\nu$ is any unit vector tangent to $M$  and $u$ is the future directed unit normal to $M$, we have along $\S$ in $M$,
\beq
\mu + J(\nu)  = G(u,\ell) = \calT(u,\ell) + \Lambda  \,.
\eeq
Thus, in the presence of ordinary matter fields one will have $\mu + J(\nu) > 0$ even if $\Lambda = 0$.  Moreover, if one assumes $\cal T$ obeys the dominant energy condition (which includes the matter vacuum case  $\mathcal{T} = 0$) and $\Lambda > 0$, then one has 
\beq\label{lambdalb}
\mu + J(\nu)  \ge \Lambda  \,,
\eeq
in which case $\mu + J(\nu)$ has a positive lower bound on all of $M$.

\proof[Proof of Proposition \ref{infrigid}] We have that $\l_1(L) \ge 0$, where $L$ is the MOTS stability operator.   Hence we may apply Lemma \ref{keyfact}.  Since $\kappa_\S = \frac12 R_\S$, 
inequality~\eqref{stabineq}, with $f =1$, implies,
\beq\label{ineqGB}
\int_\S \left( \mu + J(\nu) +\frac12 |\chi|^2\right) dA \le \int_\S \k_\S dA = 4\pi  \,.
\eeq
On the other hand, by the definition of the constant $c$,
\beq\label{ineqc}
\int_\S \left( \mu + J(\nu) +\frac12 |\chi|^2\right) dA \ge \int_S c \, dA = c A(\S) \,.
\eeq
Inequalities \eqref{ineqGB} and  \eqref{ineqc}  now imply \eqref{ineqA}.

Now assume $A(\S) = 4\pi/c$.   Then inequalities \eqref{ineqGB} and  \eqref{ineqc} combine to give,
\beq
\int_\S \left( \mu + J(\nu) +\frac12 |\chi|^2\right) dA = 4\pi  \,,
\eeq
or, equivalently,
\beq\label{equal}
\int_\S \left( (\mu + J(\nu) - c) +\frac12 |\chi|^2\right) dA = 0\,.
\eeq
Since $\mu + J(\nu) \ge c$ on $\S$, this implies that $\mu + J(\nu) \equiv  c$ and $\chi \equiv 0$.

We now have $Q = \k_\S -c$.  By Lemma \ref{keyfact}, $\l_1(L_0) \ge 0$.  But setting $f = 1$ in the right hand side of $\eqref{ray}$ gives zero, which implies that $\l_1(L_0) \le 0$.  Thus, $\l_1(L_0) = 0$ and $\phi =1$ is an associated eigenfunction, i.e. is a solution to
\beq
-\L \phi + (\k_\S-c)\f= 0 \, ,
\eeq
and hence $\k_\S = c$.\qed

\begin{rem}\label{eigen0}
{\rm Note that the proof also shows that $\l_1(L) = 0$.  Indeed, we have $0 = \l_1(L_0)  \ge \l_1(L) \ge 0$.}
\end{rem}

\noindent
{\it Dynamical horizons:}  The notion of a dynamical horizon was studied extensively in \cite{AK}.  By definition, a dynamical horizon (DH) is a spacelike hypersurface foliated by MOTS, subject to the additional requirement that along each such MOTS, one has $\th_- < 0$, i.e,  the future directed ingoing light rays are converging.  The view put forth in \cite{AK} (see also \cite{Hay2}) is that a DH should be regarded as a quasi-local version of a dynamical black hole.  The condition,  $\th_- < 0$, along each MOTS in the foliation is a physical requirement that, roughly speaking, distinguishes a DH as a black hole, rather than a white hole.   As shown in \cite{AG}, the foliation of a spacelike hypersurface by MOTS, if such a foliation exists, is unique. Moreover each such MOTS is stable, in fact, weakly outermost.  

Vaidya spacetime, which is  a spherically symmetric spacetime containing a null fluid, is a well-known example of a  black hole spacetime containing DHs; cf. \cite[Appendix A]{AK}.  
There is a canonical DH $M_{can}$ in Vaidya spacetime  which inherits the spherical symmetry.  Using the formulas in \cite[Appendix A]{AK}, one easily verifies that equality holds in \eqref{ineqA} for each MOTS in 
$M_{can}$, where $c$ is taken to be the greatest lower bound.\footnote{This can also be seen from general considerations.}   Now, there is a well-known nonuniqueness feature of DHs \cite{AG}.  In a sense that can be made precise, DHs are {\it observer dependent}.  Here `observer' should be understood as a family of spacelike hypersurfaces.  Consider a  family of spherically symmetric spacelike hypersurfaces in Vaidya spacetime, each cutting $M_{can}$ transversely in a MOTS. Smoothly perturbing this family in a nonspherically symmetric manner, will produce, in general, a nonspherically symmetric DH (see in particular the existence results in \cite{AMS1, AMS2}), in which the foliating MOTSs are no longer round, and hence do not saturate the area bound in  \eqref{ineqA}.   Thus, in response to a question raised in \cite{AG}, Proposition \ref{infrigid} provides a criterion for singling out the canonical DH in Vaidya spacetime without making explicit reference to the underlying spherical symmetry.  

\medskip
\noindent
{\it Axisymmetry:}  Suppose, in  Proposition \ref{infrigid}, one assumes that $\S$ is  axisymmetric 
in the sense of \cite{Dain1}, and hence admits a suitable rotational Killing vector field $\eta$.
  Then, if  $\S$ is axisymmetric-stable in the sense of \cite{Dain1},  the inequality \eqref{ineqA} can be refined.\footnote{Here we take the axisymmetric variation vector field $X$ in \cite{Dain1} to be $V = \phi \nu$, as in the sentences above \eqref{thder}.}   Using \cite[Lemma 1]{Dain1}, which refines the inequality  \eqref{stabineq} for  such MOTSs  and for axisymmetric functions 
$f$, one obtains in  a manner similar to the proof of Proposition~\ref{infrigid} (but where $Q$ in \eqref{Q} now acquires an additional nonnegative term) the area inequality,
\beq\label{ineqA2}
A(\S) \le \frac{4\pi}{c +\omega} \,,
\eeq
where $\omega$ is a nonnegative constant which is strictly positive if the angular momentum $J$ of $\S$  (see e.g. \cite{Dain1, Dain0, AK}) is nonzero.  The constant $\omega$ is, in the notation used here, the average value over $\S$ of the quantity $|K(\eta/|\eta|,\nu)|^2$.
Thus, while the angular momentum determines a lower bound for the area \cite{Dain1}, it also influences the upper bound.  If equality holds in \eqref{ineqA2} then, by  similar reasoning as before, one sees that $\mu +J(\nu) = c$ and $\chi = 0$.

Finally, we mention that results concerning  the infinitesimal rigidity of {\it noncompact} stable minimal MOTS have been obtained in \cite{Carlotto}.

\subsection{The splitting result}

We now establish a local initial data splitting result analogous to Theorem \ref{theo.BrayBrendleNeves}. For this purpose, we fix some notation and terminology. If $\S$ is a separating MOTS in $(M,g,K)$, let $M_+$ be the region consisting of $\S$ and the region outside of $\S$. 
Then $\S$ is weakly outermost if there is no outer trapped surfaces in $M_+$ homologous to $\S$.  $\S$ is outer area minimizing if its area is greater than or equal to the area of any surface in $M_+$ homologous to 
$\S$.

\begin{thm}\label{split}
{\it  Let $(M,g,K)$ be a 3-dimensional initial data set. Let $\Sigma\subset M$ be a separating spherical  MOTS in $M$ which is weakly outermost and outer area minimizing. 
Suppose that $\mu-|J|\ge c$ on~$M_+$ for some $c >0$.  Then,
if $A(\S) = 4\pi/c$,  the following hold.
\ben
\item  
An outer neighborhood $U\approx[0,\varepsilon)\times\Sigma$ of $\Sigma$ in $M$ is isometric to $([0,\varepsilon)\times\Sigma,dt^2+ h)$, where $h$ is the round metric on $\S$  
of radius $1/\sqrt{c}$.
\item  Each slice $\Sigma_t\approx\{t\}\times\Sigma$ is totally geodesic as a submanifold of spacetime.
Equivalently, $\chi_+(t) = \chi_-(t) = 0$.

\item  $K(\cdot,\cdot)|_{T\Sigma_t}= 0$,  $K(\nu_t,\cdot)|_{T\Sigma_t} =0$, where $\nu_t$ is the outer unit normal to $\S_t$, and $J = 0$.
\een
}
\end{thm}


\proof  As observed in Section 2, weakly outermost MOTSs are stable.  Hence, since $\mu - J(\nu) \ge \mu - |J| \ge c$ and, by assumption, $A(\S) = 4\pi/c$,  Proposition \ref{infrigid} applies.  Thus, by Remark \ref{eigen0} we have that $\l_1(L) = 0$, where $L$ is the MOTS stability operator of $\S$.

We now recall an argument from \cite{Grigid} to show that an outer neighborhood of $\S$ is foliated 
by {\it constant} null expansion hypersurfaces with respect to a suitable scaling of the
future directed outward null normals.

For $f\in C^{\infty}(\S)$, $f$ small, let $\Th(f)$ denote the null  expansion of the hypersurface $\S_f : x \to exp_x f(x) \nu$ with respect to the (suitably normalized)
future directed outward null normal field to $\S_f$.  $\Th$ has linearization, $\Th'(0) = L$.
We introduce the
operator,
\begin{align}
\Th^* : C^{\infty}(\S) \times \bbR \to  C^{\infty}(\S) \times \bbR  \,, \quad 
\Th^*(f,k) = \left(\Th(f) -k , \int_{\S}f\right)  \,.
\end{align}
Since, by Lemma \ref{prin}(i), $\l_1(L)=0$ is a simple eigenvalue, the 
kernel of $\Th'(0)=L$ consists only of constant multiples of the positive
eigenfunction $\phi$.   We note that $\l_1(L)=0$ is also a simple eigenvalue
for the adjoint $L^*$ of $L$ (with respect to the standard
$L^2$ inner product on $\S$), for which there exists
a positive eigenfunction $\phi^*$; cf. \cite{AMS2}.  Then the equation
$Lf =v$ is solvable if and only if $\int v\phi^* = 0$.
From these facts it follows easily that  
 $\Th^*$ has invertible linearization about $(0,0)$.  
Thus, by the inverse function theorem,
for $s\in \bbR$ sufficiently small there exists $f(s) \in C^{\8}(\S)$ and $k(s)\in \bbR$ 
such that,
\begin{align}\label{inverse}
\Th(f(s)) = k(s) \qquad \mbox{and} \qquad \int_{\S} f(s) dA = s\,.
\end{align}
By the chain rule, $\Th'(0)(f'(0)) = L(f'(0)) = k'(0)$.    The fact that $k'(0)$ is orthogonal
to $\phi^*$ implies  that $k'(0) = 0$.
Hence $f'(0) \in {\rm ker}\, \Th'(0)$.  The second equation in (\ref{inverse})
then implies that $f'(0) = const \cdot \f > 0$.   

It follows that 
for $s$ sufficiently small, the hypersurfaces $\S_{f_{s}}$ form a smooth foliation
of a neighborhood of $\S$ in $M$ by hypersurfaces of constant null expansion.
Thus, one can introduce coordinates $(t, x^i)$ in a neighborhood $W$ of $\S$ in $M$,
such that, with respect to these coordinates, $W = (-\vep,\vep) \times \S$, and for each
$t \in  (-\vep,\vep)$, the $t$-slice $\S_t = \{t\} \times \S$ has constant null expansion 
$\th(t)$ with respect to $\ell_t$, where $\ell_t = u+\nu_t$,
and $\nu$ is the outward unit normal
field to the $\S_t$'s in $M$.  In addition, the coordinates $(t,x^i)$ can be chosen
so that  $\frac{\d}{\d t} = \phi \nu$, 
for some positive function $\phi = \phi(t,x^i)$ on $W$.  The metric $g$ expressed with respect to these coordinates is given by
\beq\label{metric}
g|_W = \phi^2 dt^2 + h_t
\eeq
where $h_t =h_{ab}(t, {\bf x}) dx^a dx^b$ is the induced metric on $\S_t$.

Next, we want to show for $t \in [0,\vep)$,  that $\S_t$ is a MOTS, i.e., $\th(t) = 0$.  The assumption that 
$\S$ is weakly outermost, together with the constancy of $\th(t)$, implies that $\th(t) \ge 0$ for all $t \in [0,\vep)$.  We now derive the reverse inequality.  

A computation  similar to that leading to (\ref{thder}) (but where we can no longer
assume $\th$ vanishes)
shows that the null expansion function $\th = \th(t)$ of the foliation obeys the evolution equation, 
\beq\label{thder2}
\frac{d\th}{dt}  =  -\triangle \phi + 2\<X,\D\phi\> +  \left(Q  +{\rm div}\, X - |X|^2 + \th\tau -\frac12 \th^2 \right)\phi \,,
\eeq
where $\tau$ is the mean curvature of $M$.  It is to be understood in the above that, for each $t$, the terms live on $\S_t$,   e.g., $\triangle = \triangle_t$ is the Laplacian on $\S_t$, 
$Q =Q_t$ is the scalar on $\S_t$, defined as in \eqref{Q}, and so on.

Since $\phi>0$, a manipulation using (\ref{thder2})  gives,
\begin{eqnarray}\label{thph}
 \frac{\theta'}{\phi}&=&-\frac{\Delta\phi}{\phi}+2\langle X,\frac{\nabla\phi}{\phi}\rangle-|X|^2+\div X+Q
 +\theta\tau-\frac{1}{2}\theta^2 \nonumber \\
 &=&\div Y-|Y|^2+Q+\theta\tau-\frac{1}{2}\theta^2\\
 &\le&\div Y+Q+\theta\tau,\nonumber
\end{eqnarray} where $Y=X-\nabla\ln\phi$. 
Then, since $\theta'(t)$ is also constant on $\Sigma_t$,  we obtain
\begin{eqnarray*}
 \theta'\int_{\Sigma_t}\frac{1}{\phi}dA_t-\theta\int_{\Sigma_t}\tau dA_t&\le&\int_{\Sigma_t}QdA_t\\
 &=&\int_{\Sigma_t}(\kappa-(\mu+J(\nu))-\frac{1}{2}|\chi|^2)dA_t\\
 &\le&\int_{\Sigma_t}(\kappa-(\mu-|J|)-\frac{1}{2}|\chi|^2)dA_t\\
 &\le&\int_{\Sigma_t}(\kappa-c)dA_t\\
 &=&4\pi-cA(\Sigma_t)\\
 &=&cA(\Sigma)-cA(\Sigma_t)\\
 &\le&0,
\end{eqnarray*} 
where above we have used the Gauss-Bonnet theorem and the assumption that $\Sigma$ is outer area minimizing.  Thus, $\th' - \a \th \le  0 
$ where $\a(t) = \int_{\Sigma_t}\tau dA_t/\int_{\Sigma_t}\frac{1}{\phi}dA_t$, and hence,
\beq
(e^{-\int_0^t \a dt} \th)' \le 0 \quad \text{for $t \in [0,\vep)$}  \,.
\eeq
Since $\th(0) = 0$, it follows that $\th(t) \le 0$ for $t \in [0,\vep)$.

Thus, $\th(t) = 0$ for all $t \in [0,\vep)$, i.e., each $\S_t$, with $t \in [0,\vep)$, is a MOTS.  Moreover, each $\S_t$ is weakly outermost and hence stable.  Inequality \eqref{ineqA} applied to $\S_t$, together with the fact that $\S$ is outer area minimizing, implies that $A(\S_t) = 4\pi/c.$ 
It follows that for each $\S_t$,  
$t \in [0,\vep)$, $\k \equiv c$ (and hence $\S_t$ is a round sphere of radius $1/\sqrt{c}$), $\mu +J(\nu) \equiv c$, $\chi \equiv 0$, and hence $Q \equiv 0$.  Then, setting $\th \equiv 0$ in equation \eqref{thph} and integrating over $\S_t$ gives,
\beq
X = \frac{\D \phi}{\phi} \quad \text{on each } \S_t,\,  t \in [0,\vep) \,.
\eeq

Now, since $A(\Sigma_t)=A(\Sigma)$, $\Sigma$ is outer area minimizing, and $\Sigma_t\subset M_+$ is homologous to $\Sigma$, it follows that $\Sigma_t$ is also 
outer area minimizing. Then, the mean curvature $H(t)$ of $\Sigma_t$ must be nonnegative, for each $t\in[0,\varepsilon)$, otherwise $\S_t$ could be perturbed to a surface of smaller area. 
By the first variation of area, we have
\begin{eqnarray*}
 &0=\dfrac{d}{dt}A(\Sigma_t)=\displaystyle\int_{\Sigma_t}H(t)\phi dA_t,&
\end{eqnarray*} which implies $H(t)=0$ for each $t\in[0,\varepsilon)$, since $\phi>0$.

Now, because $\Sigma_t$ is a minimal MOTS, $\tr_{\Sigma_t}K=0$, for each $t\in[0,\varepsilon)$. Then, the null mean curvature $\theta_-(t)=
\tr_{\Sigma_t}K-H(t)$ of $\Sigma_t$ with respect to $l_-(t)=u-\nu_t$ also vanishes. 

Therefore, applying \eqref{thder2}    
for $\theta_-$ and $\phi_-=-\phi$ instead of $\theta=\theta_+$ and $\phi$, respectively, we have
\beq\label{thderminus}
0=\theta_-'=-\triangle\phi_-+2\langle X_-,\nabla\phi_-\rangle+(Q_--|X_-|^2+\div X_-)\phi_- \,,
\eeq
where
\begin{align}
Q_- &=\k-(\mu+J(-\nu))-\frac{1}{2}|\chi_-|^2=c-(\mu+|J|)-\frac{1}{2}|\chi_-|^2  \nonumber \\
&= - 2 |J| - \frac{1}{2}|\chi_-|^2 \,, \,\, \text{and}   \label{Qminus} \\
X_- &=  (K(-\nu_t,\cdot)|_{T\Sigma_t})^\sharp=-X = - \frac{\D \phi}{\phi}  \,. \label{Xminus}
\end{align}
Substituting \eqref{Qminus} and \eqref{Xminus} into \eqref{thderminus}, with $\phi_- = -\phi$ leads to,
\beq
\triangle \phi + \frac{|\D\phi|^2}{\phi} + \left(|J| +\frac14 |\chi_-|^2 \right) \phi = 0 \,,
\eeq
which, after integrating over $\S_t$, implies
\beq
X = \D \phi  = J = \chi_- = 0 \,\,\, \text{on } \S_t \,,\quad t \in [0,\vep)  \,.
\eeq
Equation \eqref{nullforms} now implies that $K|_{T\S_t} = 0$ and that  $(\S_t, h_t)$ is totally geodesic in $M$, for each $t \in [0,\vep)$.  It now follows that $\phi = \phi(t)$ is a function of $t$ only and that $h_t$ does not depend on $t$, $\frac{\d h_{ab}}{\d t} = 0$. With the simple change of variable 
$d s = \phi(t) dt$ in \eqref{metric}, this in turn implies that $g$ has product structure on $W_+ = W \cap M_+$,
\beq
g|_{W_+} =ds^2 + h_0 \,,
\eeq
where $h_0$ is the round metric on the sphere of radius $1/\sqrt{c}$.  This completes the proof of 
Theorem~\ref{split}.\qed


\smallskip
\begin{rem}\label{remark.3.2}
{\rm  Using that $d\tau=d(\tr K)=\div K$ (because $J=0$), $K|_{T\Sigma_t}=A=0$, and 
 $K(\nu_t,\cdot)|_{T\Sigma_t}=0$, we can see that $\tau$ does not 
 depend on $\Sigma_t$, i.e., $\tau$ depends only on $t\in[0,\varepsilon)$.}
 \end{rem}

\section{The Nariai spacetime}

The Nariai spacetime  is a simple exact solution to the  vacuum ($\mathcal{T} = 0$) Einstein equation \eqref{eineq} with positive cosmological constant $\Lambda > 0$.   It is a metric product of $2$-dimensional de Sitter space and $S^2$,
 \beq
\bar N = (\bbR \times S^1) \times S^2 \,, \quad \bar h = \frac1{\Lambda}\left(-dt^2 + \cosh^2 t \,d\th^2  + 
d\Omega^2\right)  \,.
 \eeq
 As discussed in \cite{Bousso} (see also \cite{Hay1}), the Nariai spacetime is an interesting  limit of Schwarzschild-de Sitter space, as 
the size of the black hole increases and its area approaches the upper bound in \eqref{ineqA}, with 
$c = \Lambda$.   

In this section we show that the initial data sets considered in Theorem \ref{split} can be 
realized as spacelike hypersurfaces in the Nariai spacetime.  

\begin{thm}\label{nariai}
 Let $(M,g,K)$ be a 3-dimensional initial data set. Under the same assumptions of Theorem \ref{split}, an outer neighborhood 
 $U\approx[0,\varepsilon)\times\Sigma$ of $\Sigma$ in $M$ can be  embedded into the 4-dimensional Nariai spacetime $(\bar N,\bar h)$ as 
 a spacelike hypersurface so that $g|_U$ is the induced metric from $\bar N$ and $K|_U$ is the second fundamental form of $U$ in $\bar N$.
\end{thm}

We begin with some preliminary computations.  For the sake of convenience we set $\Lambda = 1$.  Then observe that $(\bar N,\bar h)$ is locally isometric to 
\begin{eqnarray*}
 &\tilde N=\R\times\R\times S^2,\ \ \ \tilde h=-dt^2+(\cosh^2t)dr^2+d\Omega^2.&
\end{eqnarray*} 

Given a smooth function $t:I\subset\R\to\R$, consider the hypersurface 
\begin{eqnarray*}
 &N=\{(t(s),r(s),p):s\in I,p\in S^2\}\subset\tilde N,&
\end{eqnarray*} where
\begin{eqnarray*}
 &r=\displaystyle\int\frac{\sqrt{1+(t'(s))^2}}{\cosh t(s)}ds.&
\end{eqnarray*} Let $\tilde Z(t,r,x,y)=(t,r,\varphi(x,y))$ be a local parametrization of $\tilde N$, where $\varphi$ is a local parametrization of $S^2$, 
and $Z(s,x,y)=\tilde Z(t(s),r(s),x,y)$ be a local parametrization of $N$. The local coordinate vector fields $\{Z_s,Z_x,Z_y\}$ on $N$ are given by
\begin{eqnarray*}
 Z_s=t'\partial_t+r'\partial_r,\, Z_x=\partial_x,\, Z_y=\partial_y,
\end{eqnarray*} where $\{\partial_t,\partial_r,\partial_x,\partial_y\}$ are the local coordinate vector fields on $\tilde N$. If $h$ is the 
induced metric on $N$, we have
\begin{eqnarray*} 
 &h_{ss}=-(t')^2+(r')^2\cosh^2t=-(t')^2+\left(\dfrac{\sqrt{1+(t')^2}}{\cosh t}\right)^2\cosh^2t=1,&\\
 &h_{sx}=h_{sy}=0,&\\
 &h_{xx}=(d\Omega^2)_{xx},\, h_{xy}=(d\Omega^2)_{xy},\, h_{yy}=(d\Omega^2)_{yy}.&
\end{eqnarray*} Then, $N$ is a spacelike slice in $\tilde N$ isometric to the cylinder $(I\times S^2,ds^2+d\Omega^2)$.

Now, we are going to compute the second fundamental form $P$ of $N$ in $\tilde N$. Denote by $u$ the timelike future directed unit normal 
to $N$. We can see that
\begin{eqnarray*}
 &u=(1+(t')^2)^{1/2}\partial_t+\dfrac{t'}{\cosh t}\partial_r=:a\partial_t+b\partial_r.&
\end{eqnarray*}
Observe that 
\begin{eqnarray*}
 &P(Z_s,Z_x)=P(Z_s,Z_y)=P(Z_x,Z_x)=P(Z_x,Z_y)=P(Z_y,Z_y)=0.&
\end{eqnarray*} Then, the mean curvature $\sigma$ of $N$ is given by $P(Z_s,Z_s)$, which determines the second fundamental form $P$. Thus,
\begin{eqnarray*}
 \sigma&=&P(Z_s,Z_s) =  - \tilde h(u,\tilde\nabla_{Z_s}Z_s)\\
 &=&-\tilde h(a\partial_t+b\partial_r,\tilde\nabla_{Z_s}(t'\partial_t+r'\partial_r))\\
 &=&-\tilde h(a\partial_t+b\partial_r,t''\partial_t+r''\partial_r+t'\tilde\nabla_{Z_s}\partial_t+r'\tilde\nabla_{Z_s}\partial_r)\\
 &=&-[-at''+br''\cosh^2t+\tilde h(a\partial_t+b\partial_r,t'\tilde\nabla_{Z_s}\partial_t+r'\tilde\nabla_{Z_s}\partial_r)].
\end{eqnarray*} Continuing,
\begin{eqnarray*} 
 \tilde h(\partial_t,\tilde\nabla_{Z_s}\partial_t)&=&\dfrac{1}{2}Z_s\tilde h(\partial_t,\partial_t)=0,\\
 \tilde h(\partial_t,\tilde\nabla_{Z_s}\partial_r)&=&\tilde h(\partial_t,\tilde\nabla_{t'\partial_t+r'\partial_r}\partial_r)\\
 &=&\frac{t'}{2}\partial_r\tilde h(\partial_t,\partial_t)-\frac{r'}{2}\partial_t\tilde h(\partial_r,\partial_r) = -r'\sinh t\cosh t,\\
 \tilde h(\partial_r,\tilde\nabla_{Z_s}\partial_t)&=&-\tilde h(\partial_t,\tilde\nabla_{Z_s}\partial_r)=r'\sinh t\cosh t,\\
 \tilde h(\partial_r,\tilde\nabla_{Z_s}\partial_r)&=&\frac{1}{2}Z_s\tilde h(\partial_r,\partial_r)=t'\sinh t\cosh t.
\end{eqnarray*} Therefore,
\begin{eqnarray*}
 \sigma&=&-[-at''+br''\cosh^2t-a(r')^2\sinh t\cosh t+2bt'r'\sinh t\cosh t]\\
 &=&at''-br''\cosh^2t+(ar'-2bt')r'\sinh t\cosh t.
\end{eqnarray*} Observing that $r'=\frac{a}{\cosh t}$ and $a'=\frac{t't''}{a}$,
\begin{eqnarray*}
 at''-br''\cosh^2t&=&at''-\frac{t'}{\cosh t}\left(\frac{a}{\cosh t}\right)'\cosh^2t\\
 &=&at''-\left(\frac{t'}{\cosh t}\right)\frac{\dfrac{t't''}{a}\cosh t-at'\sinh t}{\cosh^2t}\cosh^2t\\
 &=&at''-\frac{(t')^2t''}{a}+a(t')^2\tanh t\\
 &=&\frac{(a^2-(t')^2)t''}{a}+a(t')^2\tanh t\\
 &=&\frac{t''}{a}+a(t')^2\tanh t.
\end{eqnarray*} Also,
\begin{eqnarray*}
 (ar'-2bt')r'\sinh t\cosh t&=&\left(a\frac{a}{\cosh t}-2\frac{t'}{\cosh t}t'\right)\frac{a}{\cosh t}\sinh t\cosh t\\
 &=&(a^2-2(t')^2)a\tanh t.
\end{eqnarray*} Finally,
\begin{eqnarray*}
 \sigma=\frac{t''}{a}+((t')^2+a^2-2(t')^2)a\tanh t=\frac{t''+a^2\tanh t}{a}, 
\end{eqnarray*} thus,
\begin{eqnarray*}
 \sigma=\frac{t''+(1+(t')^2)\tanh t}{\sqrt{1+(t')^2}}. 
\end{eqnarray*}

\proof[Proof of Theorem \ref{nariai}]  After a scaling, we can assume $c = \Lambda=1$.
 By Theorem~\ref{split}, an outer neighborhood $U\approx[0,\varepsilon)\times\Sigma$ of $\Sigma$ in $M$ is 
 isometric to the product $([0,\varepsilon)\times S^2,ds^2+d\Omega^2)$. Furthermore, by Remark \ref{remark.3.2}, 
 the mean curvature $\tau$ depends only on $s\in[0,\varepsilon)$. Then, choosing a smaller $\varepsilon>0$ if necessary, 
 we can take a solution $t:[0,\varepsilon)\to\R$ to the problem
 \begin{eqnarray*}
   \dfrac{t''+(1+(t')^2)\tanh t}{\sqrt{1+(t')^2}}=\tau,\ \ \ t(0)=t'(0)=0.
 \end{eqnarray*} 
 Then, defining $Z:[0,\varepsilon)\times\Sigma\to\tilde N$, $Z(s,p)=(t(s),r(s),p)$, where 
 \begin{eqnarray*}
  r=\int\frac{\sqrt{1+(t')^2}}{\cosh t}ds
 \end{eqnarray*}
 as before, we have an isometric embedding of 
 $U$ into $\tilde N$. Here, we are identifying $\Sigma\approx S^2$. 
 To see that $K$ is the second fundamental form of $U$ in $\tilde N$, remember that $K$ is determined by the mean curvature 
 $\tau$ $(=K(\nu,\nu))$, because $K|_{\Sigma_s}=0$ and $K(\nu_s,\cdot)|_{\Sigma_s}=0$, and $P$ is determined by the mean
 curvature $\sigma=\tau$. Then, the result follows because $(\R\times[\alpha,\beta)\times S^2,\tilde h)$ can be isometrically embedded
 into $(\bar N,\bar h)$, if $\beta-\alpha>0$ is small enough.
 
%


\providecommand{\bysame}{\leavevmode\hbox to3em{\hrulefill}\thinspace}
\providecommand{\MR}{\relax\ifhmode\unskip\space\fi MR }
\providecommand{\MRhref}[2]{%
  \href{http://www.ams.org/mathscinet-getitem?mr=#1}{#2}
}
\providecommand{\href}[2]{#2}

\end{document}